# Equation of state of SiC at extreme conditions: new insight into the interior of carbon rich exoplanets

Miozzi F.[1], Morard G.[1], Antonangeli D.[1], Clark A.N.[2], Mezouar M.[3], Dorn C.[4], Rozel A.[5] & Fiquet G.[1]

[1] Sorbonne Université, Muséum National d'Histoire Naturelle, UMR CNRS 7590, IRD, Institut de Minéralogie, de Physique des Matériaux et de Cosmochimie, IMPMC, 75005 Paris, France

[2] Department of Earth and Planetary Sciences, Northwestern University, Evanston, IL, USA, 60208

[3] European Synchrotron Radiation Facility (ESRF). Grenoble.

[4] University of Zurich, Institute of Computational Sciences, Winterthurerstrasse 109, CH-8057 Zurich

[5] Institute of Geophysics, Department of Earth Sciences, ETH Zurich, Sonneggstrasse 5, CH-8092 Zurich, Switzerland

Corresponding author: Francesca Miozzi (francesca.miozzi@upmc.fr)

Key points:

- Experimental determination of SiC thermal equation of state up to 200 GPa and 3500 K.
- We constrain the SiC B3 - B1 phase transition and expand the Si–C binary phase diagram mapping.
- Carbon enriched planets can have the same mass-radius relationship as Earth's-like planets but different dynamical behaviour.

## Abstract

There is a direct relation between the composition of a host star and that of the planets orbiting around it. As such, the recent discovery of stars with unusual chemical composition, notably enriched in carbon instead of oxygen, support the existence of exoplanets with a chemistry dominated by carbides instead of oxides. Accordingly several studies have been recently conducted on the Si–C binary system at high pressure (P) and temperature (T). Nonetheless, the properties of carbides at the P-T conditions of exoplanets interiors are still inadequately constrained, effectively hampering reliable planetary modelling. Here we present an in situ X-ray diffraction study of the Si–C binary system up to 200 GPa and 3500 K, significantly enlarging the pressure range explored by previous experimental studies. The large amount of collected data allows us to properly investigate the phase diagram and to refine the Clapeyron slope of the transition line from the zinc blende to the rock salt structure. Furthermore the pressure-volume-temperature equation of state are provided for the high-pressure phase, characterized by low compressibility and thermal expansion.

Our results are used to model idealized C-rich exoplanets of end–members composition. In particular, we derived mass-radius relations and performed numerical simulations defining rheological parameters and initial conditions which lead to onset of convection in such SiC planets. We demonstrate that if restrained to silicate–rich mantle compositions, the interpretation of mass-radius relations may underestimate the interior diversity of exoplanets.

# 1 Introduction

## 1.1 Exoplanets

Exoplanets represent one of the most vibrant current research topics in space sciences. Since the first exoplanet was discovered (e.g. Mayor & Queloz, 1995) the technological progresses, have greatly enhanced this field of research, largely fostered by the quest of Earth's twin planets. The Kepler space telescope (Boruki et al., 2011), the first telescope devoted to exoplanets discovery, detected and confirmed more than thousand exoplanets. For some of them it has been possible to measure mass and radius, and to get insights on the atmosphere's chemistry. Conversely, although fundamental for the determination of inner structure, thermal behavior, and their evolution (Unterborn et al., 2014), the composition and mineralogy of exoplanets can be only indirectly constrained. It is commonly accepted that the bulk elemental composition of planets depends on the host star composition and the condensation sequence of the accretion disk (e.g. Duffy et al., 2015, Kuchner & Seager, 2005; Mena et al., 2010; Moriarty et al., 2014). Along with sun-like stars, surveys have proved the existence of host stars with diverse chemical compositions (Bond et al., 2010 and references therein) and, in particular, with unusual high value of carbon. The carbon to oxygen ratio (C/O) deeply affects the mineralogy of a planet, defining the distributions of major elements (e.g. Fe, Si, Mg) among carbides and oxides (Bond et al., 2010). A value of the C/O ratio higher than 0.8 is expected to lead to planets that, while still involving Si, Fe and Mg, as main available elements, are prevalently made by carbon bearing compound (C-rich planets) instead of oxides (Bond et al. 2010; Kuchner & Seager, 2005). The discovery of 55 Cancri-e, a planet orbiting around a star with C/O = 1.12 ± 0.19 (Delgado Mena et al., 2010) boosted the interest for C-rich planets, promoting studies on iron and silicon carbides under high pressures. Madhusudan et al. (2012) provided the first models, suggesting that a wide range of composition in the Fe–Si–C system can satisfy the mass and radius observed for 55 Cancri-e, especially when errors on measurements are properly taken into account. A recent revision of the estimates, due to the problematic assignment of an oxygen spectral line in previous works (see details in Nissen 2013; Nissen et al., 2014; Teske at al., 2013) support a value of C/O of 0.78 (± 0.08) instead of 1.12, and more generally, revised downwards the estimates of the C/O abundances for many host. Thus, the actual existence of C-rich exoplanets, at least around the so far discovered host stars, seems unlikely. Nonetheless, in view of the increasingly large variability in stars composition, and the ongoing space survey for new exoplanets, the study of C-rich mineralogical assemblages remains of direct interest. Noteworthy, the revised C/O value for 55

Cancri-e, is still higher than the C/O of the sun (C/O = 0.58, Nissen 2013) and may lead to differences in the mineralogical assemblages of the planets. Moreover, C-rich planetesimals have been suggested to form even around moderately carbon enhanced stars (0.65 < C/O < 0.8) (Moriarty et al., 2014). In support of this, it should be noticed that silicon carbide grains (SiC) were discovered as pre-solar grains in several meteorites (e.g. Hoppe et al., 2010; Lodders & Fegley, 1995). Finally, the recent launch of the Tess mission (https://tess.gsfc.nasa.gov/) will significantly increase the number of discovered exoplanets, and very likely find candidates with exotic compositions, such as C-rich exoplanets. Therefore, investigating chemical variations in planetary systems, from Earth's like planets, to carbon enriches planets is expected to gain even further importance in the near future. . Are carbides stable at planetary conditions? Can they even be the main constituents of a planet? To address these questions we performed an experimental study under high pressure and temperature, and determined the phase diagram of the Si–C binary system and the equations of state of stable compounds, information essential to the modeling of C-rich exoplanets interior.

## 1.2 Silicon carbide

SiC has always been considered a compound of great interest for materials science, with numerous uses both at ambient and high pressure and/or temperature conditions. Among different SiC polytypes stable at ambient conditions, we focused on the 3C structure ($\beta$ – Sn), a cubic structure with a stacking made by three bilayers periodicity. This is the most studied structure, together with the hexagonal polymorph (SiC 6H), because it shows the highest symmetry and, consequently, the highest electron mobility. Pioneering works on SiC were performed with computational method (e.g. Chang & Cohen, 1987), and experimental methods, employing both static (e.g. Yoshida et al., 1993) and dynamic (e.g. Sekine & Kobayashi, 1997) compression. The bulk of carried-out work pointed out a pressure-induced phase transition to a cubic rock–salt structure, whose exact location in the P-T space is still matter of debate. Shock experiments studies locate the transition at higher pressure with respect to static experiments. According to the pioneering experimental study of Yoshida et al. (1993), in the 3C polymorph the phase transition takes place at 100 GPa under ambient temperature, while different computational studies instead indicate a pressure range between 60 and 80 GPa (Thakore et al., 2013). The latter value is also supported by the experimental work by Daviau & Lee (2017a), which located the phase transition between 60 and 70 GPa. We also note that the complete transition is not described in this study. The latest report by Kidokoro et al. (2017) is based on a rescaling of calculations made by Catti et al. (2011) and advocates for a higher transition pressure.

Pressure-volume equation of state (EoS) for the 3C structure was derived by Basset et al. (1993) and Zhuravlev et al. (2013) from Raman and Brillouin spectroscopy studies, while Nisr et al. (2017) recently employed *in situ* XRD to determine its thermal EoS. Due to its possible application as an advanced ceramic, other studies provided the thermo-physical properties of SiC (e.g. Clayton 2010; Karch 1996; Stockmeier et al., 2009) and thermal parameters of B3 SiC. However no experimental data at high temperature is available on the high-pressure polymorph with B1 structure. Wilson and Militzer (2014) first addressed the behaviour of silicon carbide at extreme conditions by a DFT (density functional theory) study. They present numerical simulations on the binary Si–C system up to 40 Mbar and provide a first model of a SiC based exoplanet. Aside from this computational study, very little is known on Si–C binary system at extreme conditions.

In this work we present the results obtained from HP – HT experiments on both the B1 and B3 polymorphs of SiC combining LH DAC and in situ XRD. The results allows us (1) to give a precise pressure range for the phase transition; (2) to define the thermal equation of state for the low and high pressure polymorphs of SiC; (3) to experimentally determine the phase diagram of the Si–C binary system up to 200 GPa, and (4) to propose mass-radius relations and dynamics for idealized C-rich putative exoplanets.

**2 Method**

**2.1 Samples**

Starting materials were synthetized via physical vapor deposition (PVD). The PVD process consists in sputtering atoms from a chemically uniform target onto a specific substrate (a glass slide in our case). Elements extraction process in the target, is enhanced by the generation of a magnetic field in a chamber filled with ionized argon plasma. The result is a homogeneous thin film deposition of the chosen chemical composition on the substrate. Starting materials synthesized with the same recipes and equipment have been employed in previous LH-DAC experiments studies (Hirose et al., 2017; Morard et al., 2017) and bring several advantages to the experiments.

The amount of material produced with one deposition is enough to provide samples for several experimental runs, guaranteeing the reproducibility of the experiments. The surface flatness also yields a better optical coupling with the IR laser and a very stable heating. PVD samples are typically made by nano-crystalline grains, which would ideally be identified with diffraction by the presence of diffuse rings on the image plate. The low atomic weight of Si and C, however, makes it difficult to clearly identify the signal coming from the starting material prior to any heating. The absence of signal on the image plate, observed for each new location on the sample, is thus a guarantee for its textural homogeneity and the pristine nature of each new heating spot.

Samples consisted in non–stoichiometric SiC deposition, with a variable thickness, from 3.5 µm to 6 µm. Three compositions were chosen, two on the C-rich side of the binary Si–C phase diagram, and one on the Si-rich side of the diagram. Electron microprobe analyses were performed on the starting materials (Centre Camparis, UPMC, Paris) using a Cameca SX100 wavelength dispersive spectrometer (WDS) to measure the two main elements (C and Si). The average of the analyses for each samples results in a composition of 72.67 at % (± 1.4) carbon and 27.33 at % (± 2.26) silicon for SiC75, 68.7 at % (± 1.47) carbon and 31.3 at % (± 1.42) silicon for SiC65, and finally 22.19 at % (± 2.71) carbon and 77.81 at % (± 2.77) silicon for SiC25.

**2.2 Cell assembly**

Samples were loaded in pre-indented rhenium gaskets (initially 200 µm thick), with a drilled hole of 120, 70 and 35 µm, for diamonds with flat culets of 250 µm and beveled culets of 150/300 and 70/300 µm diameter respectively. Dried grains of KCl were cold pressed into foils, and disks of the right diameter were cut by a femto-second laser and stored at 120°C in an oven before the loading. Samples were sandwiched between two KCl disks to provide thermal and chemical insulation from the diamonds culets. KCl is not only a suitable pressure transmitting media and thermal insulator, but also a reliable pressure calibrant. Such assemblages were loaded in membrane-driven Le Toullec-type diamond anvil cells, equipped with tungsten carbide seats designed for diamonds with a conical support (Boehler & De Hantsetters, 2004), providing a maximum 2–theta angle of 70°.

**2.3 X-ray diffraction measurements at high pressure and high temperature**

Angle dispersive x-ray diffraction experiments were performed on the high-pressure beamline ID27 at the European Synchrotron Radiation Facility (ESRF) in Grenoble (Mezouar et al., 2005). The beamline set up provides a monochromatic beam of 33KeV ($\lambda$ = 0.3738 Å, Iodine k-edge) focused on a sample area of less than 3 x 3 µm$^2$ (FWHM). Exposure time was usually between 10 and 30 s and the diffraction signal was collected on a MAR CCD detector. Heating is provided by two continuous Nd:YAG fiber laser (TEM 00) focused over an heating spot > 20 µm in diameter. The power on the two sides was adjusted independently, in order to minimize axial temperature gradients. Temperature was measured on both sides of the sample before and after the X-ray exposure (only upstream during the exposure), at the center of the hot spot region. The analysis of the thermal emission is made on a 2 x 2 µm$^2$ area selected by a pinhole placed at the entrance of the spectrometer. Temperatures are obtained by the spectroradiometric method, using reflective collective optics (Shultz et al., 2005). Uncertainties on the T measurements come from the temperature gradient in the sample and the uncertainties on the fitting procedure. Following Schultz et al., (2005), the gradient in the radial direction is less than 50 K in the X-ray spot, in virtue of the > 20 µm heating spot versus the less than 3 µm diameter spot probed by the x-ray beam. The double side laser heating and the controlled geometry of the assembly contribute to maintain the axial gradient below 100 K. Therefore, for our experiments we assumed an error on the sample's temperature of ± 150 K (Morard et al., 2011). Pressure was determined by the thermal equation of state of KCl by Dewaele et al. (2012), assuming the temperature of KCl to be the average between the temperature on the culet of the diamond (295K) and the temperature measured at the surface of the sample (Campbell et al. 2009). Uncertainties on pressure determination are obtained, as described in the supporting information (Text S1), accounting for the uncertainties on the temperature calibration for KCl.

Different experimental protocols have been used for the initial determination of the phase diagram, to refine the transition pressure and to measure the EoS. In the first case, the cells were first pressurized to a target pressure. IR lasers were then aligned and the power increased step by step. Diffractions were collected for each temperature step. Once the pressure range of the phase transition was bracketed, two additional experimental runs were performed to narrow the pressure interval. In these runs, compression was cycled, both increasing and decreasing pressure at constant temperature. The first appearance upon compression of the high-pressure polymorph peaks (or disappearance of the low pressure polymorph) was taken as a reference for the phase transition. Coexistence of the two phases was also observed in several diffraction patterns.

The software DIOPTAS (Prescher & Prakapenka, 2015) was used to integrate the 2D diffraction images. Diffraction of a CeO$_2$ powder standard allowed us to calibrate the detector distance and orientation parameters. Parameters of the unit cell were then calculated by fitting the peak positions with a pseudo–Voigt Gaussian curve, with the program PD Indexer (http://pmsl.planet.sci.kobe-u.ac.jp/~seto/).

**2.4 Equations of state**

There are several formalisms linking the variation of volume to that of pressure (isothermal equation of state), each with its own assumption and mathematical development (e.g. Angel 2000; Angel et al., 2014, 2017; Duffy & Wang, 1993; Kroll et al. 2012). Here we employed a Birch-Murnaghan equation of state, that is a "finite strain EoS", in which the strain energy of a solid undergoing compression is expressed as a Taylor series in the finite Eulerian strain $f_E$, where $f_E = \frac{1}{2}\left[\left(\frac{V_0}{V}\right)^{\frac{2}{3}} - 1\right]$ (Angel, 2000). The expansion to the third order in the finite strain yield to the relation:

$$P = \frac{3}{2} K_0 \left[\left(\frac{V_0}{V}\right)^{\frac{7}{3}} - \left(\frac{V_0}{V}\right)^{\frac{5}{3}}\right]\left\{1 + \frac{3}{4}\left[K_0' - 4\right]\left[\left(\frac{V_0}{V}\right)^{\frac{2}{3}} - 1\right]\right\} \qquad (1)$$

V is the volume, $V_0$ and $K_0$ are respectively the volume and bulk modulus at ambient pressure and reference temperature (here 300 K) and $K_0'$ the first pressure derivative of the bulk modulus ($\partial K/\partial P$). A second formalism largely used for equation of state of solids is the one by Vinet, derived from interatomic potentials and designed to better represent materials under very high compression (Vinet et al., 1986, 1987), which gives:

$$P = 3K_0 \frac{(1-f_V)}{f_V^2} \exp\left(\frac{3}{2}(K_0'-1)(1-f_V)\right) \qquad (2)$$

where $f_V = \left(\frac{V}{V_0}\right)^{\frac{1}{3}}$.

As those equations of state are both considered to satisfactorily describe the compression of solid materials over the pressure range of our interest, we decided to fit our data with both, and then choose the one giving less residuals.

The thermal pressure EoS model, in which the volumes explicitly and independently depends on both P and T, can describe the high temperature behaviour of compressed materials. This approach requires the determination of the pressure at reference T condition, i.e. 300 K isotherm in our case P(V,300K), and then thermal pressure with increasing temperature along isochors $P_{th}(V,T)$. Consequently the pressure at a given temperature is defined as:

$$P(V,T) = P(V,300K) + P_{th}(V,T) \qquad (3)$$

To calculate the second term, two main thermal pressure EoS are used in literature, the Mie Grüneisen Debye and the Thermal Pressure model (Kroll et al., 2012). The Mie Grüneisen Debye (MGD) model relies on a Debye-like approach, where the thermal-induced lattice vibrations are modelled as a series of harmonic oscillators. The vibrational energy of the lattice is expressed as a function of the temperature T normalized to the characteristic Debye temperature $\theta_D$ and written as:

$$E(T,\theta_D) = 9nRT \left(\frac{T}{\theta_D}\right)^3 \int_0^{\frac{\theta_D}{T}} \frac{t^3 dt}{e^t - 1} \tag{4}$$

where n is the number of atoms per formula unit, R the gas constant. The Debye temperature

$$\theta_D = \theta_0 \exp\left[\frac{(\gamma_0 - \gamma(V))}{q}\right] \tag{5}$$

is a function of the Grüneisen parameter (γ) given by

$$\gamma(V) = \gamma_0 \left(\frac{V}{V_0}\right)^q \tag{6}$$

where q corresponds to the logarithmic volume dependence of γ(V). γ and $\theta_D$ only depend on the volume and are constant along isochors. The vibrational energy, together with the Grüneisen parameter and the volume, define the thermal pressure, with the Ec and Tc being energy and temperature at reference conditions:

$$P_{th} = \frac{\gamma(V)}{V} [E(T,\theta_D) - E_c(T_c, \theta_D)] \tag{7}$$

In summary, in the MGD model, parameters that control the thermal pressure, and hence allow the description of the thermal state of the system, are only three: the Debye temperature, the Grüneisen parameter, and *q* (see Angel et al., 2017 and references therein).

The second approach is the Thermal Pressure (TP) model. In this model the slope of an isochor in the P-T space is written as·

$$\left(\frac{\partial P}{\partial T}\right)_V = \alpha_V K_0 \tag{8}$$

(Andersen 1984; Angel et al., 2017; Poirier, 1991) with $\alpha_V$ being the thermal expansion. Consequently, during an isochoric heating thermal pressure can be written as:

$$P_{th} = \int_{T_0}^{T} \alpha_V K_0 \partial T \tag{9}$$

The product $\alpha_V K_0$ is usually assumed to be constant along an isochor, as isochors are linear for T above the Debye temperature. Holland & Powell, (2011), also defined an approximation for the integral of Pth, explicitly introducing the dependence upon the Einstein function to account for the decrease to 0 as temperature decreases to 0 K (Angel et al., 2017, Holland & Powell 2011).

$$\xi = \frac{u^2 e^u}{(e^u - 1)^2} \tag{10}$$

with $u = \theta_E/T$ and $\theta_E$ being the Einstein temperature. Following Holland and Powell (2011) $P_{th}$ can then be written as:

$$P_{th} = \alpha_{V0} K_0 \left(\frac{\theta_E}{\xi_0}\right)\left(\frac{1}{\exp\left(\frac{\theta_E}{T}\right)-1} - \frac{1}{\exp\left(\frac{\theta_E}{T_0}\right)-1}\right) \tag{11}$$

with $\xi(T_0)$ as in the thermal expansion equation from Kroll et al. (2012).

$$\xi(T_0) = \frac{\left(\frac{\theta_E}{T_0}\right)^2 \exp\left(\frac{\theta_E}{T_0}\right)}{\left[\exp\left(\frac{\theta_E}{T_0}\right)-1\right]^2} \tag{12}$$

The Einstein temperature ($\theta_E$) can be assessed from the measured entropy. In particular, for end-members the value of $\theta_E = 10363/(S/n + 6.44)$, with S being the molar entropy and n the numbers of atoms (Holland & Powell, 2011). Alternatively, the Einstein temperature can be calculated from the Debye temperature, employing the relation ($\theta_E = \theta_D * 0.806$). In the Thermal Pressure model, thermal pressure only depends on two parameters: the coefficient of thermal expansion ($\alpha_V$) and the Einstein temperature ($\theta_E$) (which is usually fixed).

Both the MGD and the TP thermal model were used to fit our data. Indeed MGD is one of the most common parameterization to represent the P-V-T relations of a solid material under extreme conditions, and has already been used for the B3 structure by Nisr et al. (2017). On the other hand, TP is typically used for rigid structures and recent studies (e.g. Angel et al., 2017, Milani et al., 2017) demonstrate that TP provides a reliable and trustworthy thermal model especially for compounds that, as SiC, have properties similar to diamond. However, it should be noted that assuming the product $\alpha_V K_0$ constant along an isochor is an approximation and, as such, it might not be always adequate (Jackson & Ridgen, 1996). A discussion about the results and which model better matches our experimental data set will be presented in the EoS section.

# 3 Results and discussion

## 3.1 Phase transition

High-pressure and high-temperature experiments were performed up to 205 GPa and 3500 K. As the starting materials were non-stoichiometric compounds, heated samples systematically disproportionated into SiC + Si or SiC + C, thus allowing us to study stoichiometric SiC simultaneously with the Si–C phase relations. KCl peaks only, were visible on the diffraction patterns collected before laser heating. As previously said, with nanocrystalline grains produced by the PVD process, the low weight of both Si and C makes extremely difficult a clear identification of the diffuse signal on the image plate before any heating. SiC diffraction peaks indeed start being detectable after initial heating and recrystallization during lasers alignment. The B3 SiC diffraction pattern is then characterized by four reflections, 111, 200, 220 and, at higher angle, 311, sometimes merged with the 211 reflection from KCl. At low pressure, the most intense reflection is the 111 (Figure 1 and Supporting Information). After the phase transition to the B1 structure, occurring between 65 – 70 GPa, the 111 reflection shifts closer to the 110 reflection of KCl, almost merging with it, and the 200 becomes the most intense reflection. As highlighted in Figure 2a the phase transition involves a significant volume reduction, with the value of the cell parameters changing, depending from actual T, from a = 4.09 – 4.10 Å ($\pm$ 0.002) (V= 68.80 – 69.06 ($\pm$ 0.1) Å$^3$) to $a$ = 3.85 – 3.88 ($\pm$ 0.004) Å (V = 57.4 – 58.6 ($\pm$ 0.2) Å$^3$). Our measurements are in agreement with previous reports in literature (e.g. Daviau & Lee 2017a, Thakore et al., 2013, Yoshida et al., 1993) and advocate for a structural transition from a cubic zinc-blende structure (B3) to a cubic rock-salt structure (B1), driven by a coordination change from tetrahedral to octahedral (Figure 2b).

The large P–T range explored during our experiments and the large number of data points allowed us to narrow down the range for the B3 – B1 transition of SiC compound (Figure 3), solving the controversy between previous studies (e.g. Daviau & Lee 2017a, Yoshida et al., 1993). Noteworthy, the fine texture of our samples, highlighted by the continuous diffraction rings on our collected patterns, helps to unambiguously define the structural transition. Our findings clearly place the transition at 66 $\pm$ 2 GPa at room temperature and between 65 and 70 GPa at high T, in agreement with suggestion from previous LH DAC experiments (Daviau & Lee 2017a) and the computational work (Thakore et al., 2013). This transition pressure is significantly lower than earlier determination at ambient temperature by Yoshida et al., (1993) and the recent proposition at high temperature by Kidokoro et al., (2017). As in Daviau and Lee (2017a), we find no evidence of intermediate structures on the transition pathway, such as the orthorhombic or tetragonal structures reported in several computational works (e.g. Catti 2001; Durandurdu et al., 2004).

Diffraction patterns showing the coexistence of both phases define a steep, almost vertical (0.001 GPa/K) Clapeyron slope in the P–T space for the B3 – B1 phase transition (Figure 3). This behavior differs from what proposed by Daviau and Lee (2017a). Indeed, if their first point at higher pressure showing coexistence falls within our coexistence range, their second point is located in a region where we only observe the stable B3 structure. Comparison with Kidokoro et al. (2017) further supports a negative Clapeyron slope, although, as already mentioned, there is a shift in the reported pressure range of the phase transition (which is 10 GPa lower in our case).

### 3.2 Equation of state

### 3.2.1 Room temperature EoS

The room temperature equation of state and the thermal equation of state were defined for both the low-pressure B3 and the high-pressure B1 structures. The data taken at ambient temperature were fitted with both the 3$^{rd}$ order Birch Murnaghan (BM3) and the Vinet equation of state. As laser heating was required to crystallize the sample, it was not possible to collect data points below 25 GPa where melting of the KCl pressure medium occurs at too low temperature. Thus the lack of those points does not allow a reliable determination of $V_0$. As a consequence a $V_0$ value fixed to that determined by Nisr et al. (2017) was employed to fit the B3 structure EoS. The fitting was performed using the EosFit suite (Angel et al., 2014; Gonzales-Platas et al., 2016). Obtained fit parameters, along with the data from previous work, are summarized in Table 1. Parameters obtained using BM3 and Vinet are within the error bars, but we observe a better fit using BM3 (for the rock-salt structure $\chi^2$ are 5.1 and 5.43 respectively for BM3 and Vinet). Accordingly, we decided to employ the values obtained with the BM3 equation of state as a reference for the thermal EoS.

Our compressibility value for the B3 structure is lower than the values found in previous XRD work on powders (e.g. Nisr et al., 2017; Yoshida et al., 1993) but is comparable with the values reported for SiC single crystal using combined XRD and Brillouin spectroscopy (Zhuravlev et al. 2013). Regarding the B1 structure, the obtained values for $K_0$ and $K_0$' are respectively higher and lower than those proposed by Kidokoro et al. (2017). The differences can be partially explained by the different orders used for the Birch Murnaghan equation of state (second order for Kidokoro et al. (2017), i.e. $K_0$' fixed to 4, and third order in this study). However, we consistently obtain a higher bulk modulus of 323-324 GPa, even when the reference volume is fixed. It is thus unlikely this difference only originates in the pressure scale. We think the larger dataset and pressure range used in our study may help to set harder constraints on the EoS as it shows the need for a third order EoS.

**3.2.2 Thermal Equation of state**

The thermal equations of state were also determined for both the low pressure and high pressure structures. For the zinc-blende B3 structure, as far as Debye temperature, $\gamma_0$ and $V_0$ we used the values determined from previous literature works (i.e. Clayton 2010, Karch et al., 1996 Stockmeier et al, 2009). For the high pressure B1 structure, instead, we fixed the Debye temperature as in Varshney et al. (2015) and refined all the other parameters.

For both structures, the fitting of the equation of state was performed using the EoSFit suite (Angel et al., 2014; Gonzales-Platas et al., 2016). For the zinc blende structure we collected 105 data points with P ranging between 25 and 65 GPa and T between 300 and 2500 K. The fit of the MGD model results in a compressibility of 228 ± 7 GPa, and a $K_0$' of 3.9 ± 0.3. The only parameter refined for the thermal model (*q*) has a value of -1.3. A negative value of *q* implies that the Grüneisen parameter increases with compression. Although quite an unusual behavior, this was already reported by Nisr et al. (2017) for one of the pressure scales they used. Most likely, the negative value of q does not have a physical meaning, and simply is a numerical consequence of a fit performed over relatively limited pressure range with points showing some scatter, in particular at high temperature.

Fit to the data with the TP model, yields better residuals in pressure (*i.e.* smaller difference between actual measurements and pressure given by the EoS) (Figure 4b). The value of the Einstein temperature was fixed to that calculated from the Debye temperature of the MGD model ($\theta_E = \theta_D * 0.806$). The obtained parameters for the zinc – blende B3 structure are $K_0 = 224 \pm 2$ GPa, $K_0$' = 4.08 ± 0.2 with $\alpha_V = 0.62 \cdot 10^{-5} \pm 0.1$ K$^{-1}$. The latter is lower than previous experimental determination by Nisr et al. (2017), but is close the values derived from calculations by Varshney et al. (2015). The compressibility and pressure derivative have the same values as in the Birch – Murnaghan 300 K EoS- with smaller errors on the parameters, while the values found with the thermal MGD model are slightly different. The isotherms shown in Figure 4a nicely account for the entire data set with a $\chi^2$ of 0.70.

In the stability field of the high-pressure rock-salt (B1) structure 348 points were used to refine the thermal equation of state, over a pressure range between 65 and 205 GPa, and temperatures between 300 and 3000 K. In the MGD thermal model refinement, the Debye temperature was kept fixed to the value reported by Varshney and coauthors (2015) and the reference pressure is considered equal to zero. As a standard procedure for the fit, we first refined only $\gamma_0$ and q using the $V_0$, $K_0$ and $K_0$' obtained with the EoS at 300 K. Then, for the final processing cycle, also the value of $V_0$, $K_0$, and $K_0$' were refined together with the other parameters of the thermal model. The best-fit solution yields $V_0 = 65.90 \pm 0.04$ Å$^3$ $K_0 = 340 \pm 10$ GPa and $K_0$' = $3.06 \pm 0.06$, with a Debye temperature of 1200 K, $\gamma_0 = 0.50 \pm 0.05$ and q = $1.6 \pm 0.5$, with $\chi^2 = 2.46$. The here-obtained value of $\gamma_0$ is lower (0.5 vs. 1.06) than that from previous computational work (i.e. Varshney et al., 2015). Effect of the dependence on $\gamma_0$ of the fits was also investigated (details in Supporting Information, Text S2). The plot of the pressure residuals shown in Figure 5b highlights the high quality of the fit, which stays within $\pm$ 5 GPa over the whole P-T range investigated.

The same procedure, fixing the value of the Einstein temperature, was followed for the TP model. The obtained results are $V_0 = 65.9 \pm 0.04$ Å$^3$, $K_0 = 339 \pm 2$ GPa, $K_0$' = $3.03 \pm 0.02$ and $\alpha = 0.58 \times 10^{-5} \pm 0.01$ K$^{-1}$ with the Einstein temperature calculated from of the Debye temperature of B1 structure. The refined values for $V_0$, $K_0$ and $K_0$' are the same as for the MGD model within error bars, and $\chi^2 = 2.67$ (see Figure S6). The coefficient of thermal expansion fall within the range of values derived from theoretical models (Varshney et al. 2015) which also predicted a significant decrease across the transition from the B3 to the B1 structure. The obtained results are presented in Table 2. Refined $V_0$, $K_0$ and $K_0$' show a reasonable agreement, i.e. almost within error bars, with parameters determined at ambient temperature. Accordingly, we consider the $V_0$, $K_0$ and $K_0$' obtained with the two thermal models as our preferred set of parameters to describe the HP-HT behavior of SiC high pressure (B1) structure.

**3.3 Phase diagram**

Having performed measurements with starting materials of different composition allowed us to probe not only SiC compound EoS but also to explore the binary Si–C system. Notably The three different Si–C compositions employed for the experiments enabled us to study both the silicon rich side and the carbon rich side of the phase diagram. Our main conclusion regarding the binary mixing is that, over the entire pressure and temperature range of this study, no intermediate compound has been observed. A SiC + Si assemblage is stable on the Si-rich side of the diagram, whereas a SiC + C assemblage is observed on the C-rich side of the diagram (Figure 6). This is in disagreement with the stability of silicon dicarbide ($SiC_2$) predicted by theory at pressure of 25 GPa by Andrew et al. (2012) and suggested to appear only above 10 Mbar by Wilson and Militzer (2014). In addition, we do not observe any decomposition of SiC compound or change in diffraction peaks intensity over 2000 K, in the pressure range 30-60 GPa as described in Daviau and Lee (2017b).

On the Si-rich side of the diagram, we have been able to determine the melting temperature at 2100 K and 60 GPa for the sample SiC25, based on the disappearance of the solid diffraction peaks and a simultaneous plateau in the laser power versus temperature curve. Due to the low Z of the material, we could not detect the appearance of diffuse scattering. Our melting criterion, thus possibly leads to overestimating of the melting temperature (Figure 6). As no other signatures of melting were observed, we hypothesize the composition with 77.81 at % Si, 22.19 at % C to be close to the eutectic. On the C-rich side no melting was detected in the whole examined pressure range for T up to 3500 K. This could be related to the difference of melting temperature of the two end members, Si and C. Diamond has indeed a very high melting temperature, much larger than that of Si. Such a large difference in the melting temperature can have strong implications for planetary dynamics through melt production for any small Si enrichment, whereas the two phases assemblage (i.e. SiC + C) can stay solid in C-rich systems. Overall, the resulting phase diagram is in agreement with the phase diagrams from Wilson & Militzer (2014).

**4 Implications for exoplanets studies**

**4.1 Mass-radius relation for SiC based planets**

The parameters obtained on the Si–C binary system for the equations of state and phase diagram, were used to model mass-radius curves for different C-rich planets composition (Figure 7). Extrapolations have been limited by the validity of the used equations. Accordingly we only modeled planets up to 5 times the mass of the Earth. Similar models have already been proposed in the past, usually plotting mass-radius-relations for archetypal planets made entirely by one element (i.e. carbon or iron) or one compound (i.e. $MgSiO_3$, SiC, $H_2O$), as in Duffy et al. (2015).

In this study, along with the mass-radius plots for commonly considered end members (i.e. pure SiC, $MgSiO_3$), we present the mass-radius curves for several idealized carbon-rich differentiated planets in which a pure iron core is assumed. The proposed inner compositions represent end-members that are unlikely to occur in nature, due to the absence of both Mg and O, as equally unlikely end-members (due to the lack of major rock-forming elements) are planets entirely made of pure $MgSiO_3$ or pure Fe. Nonetheless, such Mg- and O- free end-members were already studied by different authors (e.g. Madhusudan et al., 2012; Nisr et al., 2017; Wilson & Militzer, 2014). Studying end-member compositions, we provide reference limit values of the bulk density that can be used to interpret the possible interiors of such exoplanets. Planets less dense than pure $MgSiO_3$ for example, are classically believed to incorporate a high amount of volatiles. Very interestingly, our data show that planets with carbon-rich interiors can also be less dense than a pure $MgSiO_3$ (see Figure 7).

Along with a hypothetical planet made by pure SiC, as already done in previous literature, we also considered two models that include Fe, in different proportions, as Fe it is consider one of the most common elements in C-rich exoplanets. In the first model of an Earth-like planet, the iron core makes up a third of the planets mass. In the second model, the proportions of Fe and SiC are chosen so as the bulk composition of the planet matches solar Fe/Si abundances (solar Fe/Si) (Lodders et al., 2009). In the pressure range examined in this experimental study, our density estimates agree very well with theoretical DFT calculations from Wilson & Militzer (2014) (i.e., less than 1% variation on R at 2 Me for pure SiC planets). As expected, the addition of iron to a pure SiC interior increases the planet density. The Fe + SiC model with core mass proportions similar to Earth's core yields to a mass-radius relationship very similar to the Earth-like case. Thus, in this case, it is almost impossible to discriminate between an Earth-like interior and a carbon-rich interior on the sole basis of mass and radius. This highlights the importance of having independent constrains on the composition of planetary interiors.

**4.2 Estimations on the geodynamics of SiC planets**

The thermal evolution of a planet is governed by its geodynamical regime, affecting both deep and shallow processes. The growing knowledge about properties of Earth's minerals contributes to improve the models representing the dynamical behavior of silicate-rich planets. Nevertheless little is known about the potential effects of carbides, in particular SiC, on convection.

Studies focused on deep Earth's minerals highlighted how phase transitions might have a first order effect on the dynamics of a planet. The upper to lower mantle transition at 660 km depth is known to have an effect on the dynamic topography (Flament et al., 2013) and the geoid (Hager et al., 1984; Ricard et al., 1993), due to an increase in viscosity (~30). On the other hand, phase transitions such as the perovskite to bridgmandite generate a small viscosity jump (<<1) (e.g. Nakagawa & Tackley 2011; Yamazaki et al., 2006). As SiC undergoes a phase transition that involves a large volume change, it is of primary interest to evaluate the effect of a potential viscosity jump on the rheological behavior and consequently, on planet's dynamics. Unfortunately, parameters such as viscosity, deformation mechanism and activation energies have only been evaluated at ambient pressure on SiC compounds often doped with other chemical elements, to enhance the quality for industrial application. The absence of those parameters for the pure compound, coupled with the uncertainties regarding the characteristic of C-rich exoplanets thus limits the possibilities of running complex numerical models. We thus decided to investigate how different values of viscosity jump and activation volume affect the dynamic of a planet, by mapping the onset of mantle convection, for different initial temperatures.

We performed numerical simulations on a pure SiC mantle for a planet with 1 Earth mass (B3–B1 phase transition occurring around 1830 km depth), employing the thermodynamic parameters obtained in this study and the rheological data from Carter et al. (1984). All details about the employed equations, numerical set up and boundary conditions are provided supporting information (Text S4). A surface thermal conductivity of 270 W/m/K (Goldberg et al., 2011) was employed in this study. To account for high pressure effects, the conductivity across the mantle was multiplied by the ratio between density under pressure and at the surface.

The obtained results are summarized in Figure 8. The viscosity jump from 0.1 to 100 was imposed at the phase transition and we explored values of the activation volume from $2 \cdot 10^{-6}$ to $6 \cdot 10^{-6}$ m$^3$/mol together with three different initial potential mantle temperatures: 1900, 2100, and 2300 K. For each initial temperature our results show that convection only starts in simulations considering low activation volumes ($3 \cdot 10^{-6}$ to $6 \cdot 10^{-6}$ m$^3$/mol). The viscosity jump employed seems to have little impact on convection as the boundary curves are quite vertical. The boundary between simulations undergoing convection and the stable ones can be approximated by the following equation

$$V_c = 3.62 \cdot 10^{-6} + \frac{(T_i - 1900)^{1.205}}{400} 6.9 \cdot 10^{-7} - 3.3 \cdot 10^{-7} \log(\Delta\eta) \qquad 13$$

where $V_c$ is the critical activation volume below which convection starts. This boundary coincides with the blue, purple and red thick lines in Figure 8 (i.e. for initial temperature equal to 1900, 2100 and 2300 K).

In conclusion we observed that the onset of convection (after the magma ocean crystallization) strongly depends on the activation volume and the initial temperature. Activation volumes required for convection (~ 3 – 6 $10^{-6}$ m$^3$/mol) have been found to be lower by a factor of 2 to 10 than typical values for Earth materials. Our estimations demonstrate that dynamics in C-rich exoplanets might be far from what is commonly accepted for Earth. Moreover, we provide a first exploration of the possible geodynamic scenario for planets made by pure SiC.

**4 Conclusions**

Following the renewed interest for carbon bearing species and driven by the potential discovery of C-rich exoplanets, our experimental work provides useful information on the behavior of SiC, often considered as one of the main component of carbon-rich exoplanets. In this study we narrow the pressure range for the phase transition. We report a steep Clapeyron slope (0.001 GPa/K) with a precise determination of the volume and density changes across the B3–B1 transition of this compound. We also provide new thermal equations of state, extended up to 200 GPa and 3000 K. Furthermore, our data can be used to model M/R plots and infer dynamics for different types of C-rich exoplanets. Our results shows that it is difficult to distinguish a carbon-rich planet from an Earth-like planet solely based on bulk density and observed M/R relations. Though the improved and enhanced information we gathered on the high pressure and high temperature behavior of the possible main constituent (SiC) is crucial to narrow the range of mineralogical composition likely to be present in such C-rich exoplanets.

**Acknowledgments**

The authors wish to thank Stany Bauchau and Jeroen Jacobs (ESRF) for their help with the X-ray experiments and diamond anvil cell preparation. We would like also to thank the two anonymous reviewers for their detailed comments and suggestions for improving the manuscript. Femtosecond laser micro-machining at the Institut de Minéralogie de Physique des Matériaux et de Cosmochimie (IMPMC), Paris has been developed and realized by the "Cellule Project" with the financial support of ANR 2010-JCJC-604-01.

This project and F.M., G. M. and G. F. have received funding from the European Research Council (ERC) under the European Union's Horizon 2020 research and innovation Programme (grant agreement No 670787). C.D. is funded by the Swiss National Science Foundation under the Ambizione grant PZ00P2_174028. A.B.R received founding from the European Research Council under the European Union Seventh Framework Programme (FP/20072013)/ERC grant

Table 1: P – V Equations of state fit results for both the low pressure (B3) and high pressure (B1) structure. Data from previous experimental works are also shown.

|  | Zinc blende structure | | | | | Rock salt structure | | |
| --- | --- | --- | --- | --- | --- | --- | --- | --- |
|  | This study | | Yoshida et al. 1993 | Zhuravlev et al. 2013 | Nisr et al. 2017 | This study | | Kidokoro et al. 2017 |
| EoS | BM | Vinet | BM | BM | Vinet | BM | Vinet | BM |
| $V_0$ (Å$^3$) | 82.80* | 82.80* | 82.92 | 82.96 | 82.80 | 66.3 (0.1) | 66.3 (1.4) | 67.5 |
| $K_0$ (GPa) | 224 (5) | 222 (5) | 260 | 218 | 243 | 323 (3) | 324 (2) | 235 |
| $K_0$' | 4.1 (0.3) | 4.3 (0.4) | 2.9 | 3.75 | 2.68 | 3.1 (0.43) | 3.3 (0.52) | 4 |

*Value from Nisr et al. (2017)

Table 2: Values of the thermal equation of state for the two structures, fitted with both the Mie Grüneisen Debye

| | Zinc blende structure | | Rock salt structure | |
|---|---|---|---|---|
| | MGD | TP | MGD | TP |
| Number of data points | 105 | 105 | 348 | |
| $V_0$ (Å$^3$) | 82.80* | 82.80* | 65.8 (0.04) | 65.9 (0.04) |
| $K_0$ (GPa) | 228 (7) | 224 (2) | 339 (10) | 339 (2) |
| $K_0'$ | 3.9 (0.3) | 4.1 (0.14) | 3.06 (0.06) | 3.03 (0.02) |
| T Debye (K) | 1200* | | 1200* | |
| Gamma $_0$ | 1.06* | | 0.50 (0.05) | |
| $q$ | -1.3 (0.5) | | 1.67 (0.5) | |
| T Einstein (K) | | 976.2* | | 976.2* |
| Alpha(10$^{-5}$ K$^{-1}$) | | 0.62 (0.11) | | 0.56 (0.01) |

and the Thermal Pressure models. Parameters kept fixt are marked with an asterisk.

Figure 1: Diffraction patterns collected on SiC75 along selected P–T paths (displayed in the insets) showing the phase transition from the zinc-blende structure to the rock-salt structure upon compression (left) and decompression (right). Reflections for SiC B3, B1, C and KCl are labelled. The star signs identify peaks coming from a spurious spot on the image plate.

Figure 2: a) Plot of the volume as a function of pressure at T = 300 K. Symbols represent the experimental data from this study and the lines represent different fitted EoS. Two models from this study and three available from literature; b) Structure of the two SiC polymorphs. In light pink the coordination of the atoms and in light blue the planes responsible for the most intense reflection respectively, the 111 for SiC B3 and the 200 for SiC B1.

Figure 3: Squares: data from this study at P = 40 – 90 GPa and T = 300 – 3500 K. SiC B3, B1 and B3-B1 coexistence are reported in different colors; the phase transition region is highlighted by the light red band. Diamonds illustrate the data from Daviau and Lee, (2017). The green dashed line is the slope showed in Kidokoro et al. (2017) obtained after a rescaling of results by Catti et al. (2001).

Figure 4: Results obtained fitting the zinc blende structure with the Thermal Pressure model: a) experimental data points and isotherms; b) Pressure residuals of the fit. On the right the temperature scale.

Figure 5: Results obtained fitting our data for the B1 structure with the MGD thermal model: a) experimental points and isotherms from 300 to 3000K; b) pressure residuals for the fit. On the right the temperature scale.

Figure 6: Phase diagram of the Si–C binary system at P = 60 GPa.

Figure 7: Mass-radius-relations for different idealized exoplanet interiors together with the standard comparison curves (pure Fe, $MgSiO_3$, and Earth like). We propose several C-rich end members: a pure SiC planet, a iron core + SiC mantle planet with mass proportions similar to the Earth's core and mantle, and finally an iron core + SiC mantle planet with a bulk ratio of Fe/Si that matches solar abundances. The black dots show the measured masses and radii for solar system planets and some observed exoplanets. Details on the calculation of the M/R plot are provided in the supporting information (Text S3).

Figure 8: Map of the onset of convection in all numerical simulations. Open circles show simulation in which convection started and crosses show when convection never started. Blue, purple, and red data represent cases with initial mantle potential temperature of 1900, 2100 and 2300 K, respectively. Thick lines represent the regime boundary between convection and diffusion (see equation 13). The 2D images show two examples of temperature field with convection (bottom-left) or not (top-right) in our computational domain. The isocontours show viscosities of $\eta = 10^{19}$ Pa.s (red), $\eta = 10^{21}$ Pa.s (green) and $\eta = 10^{23}$ Pa.s (blue).

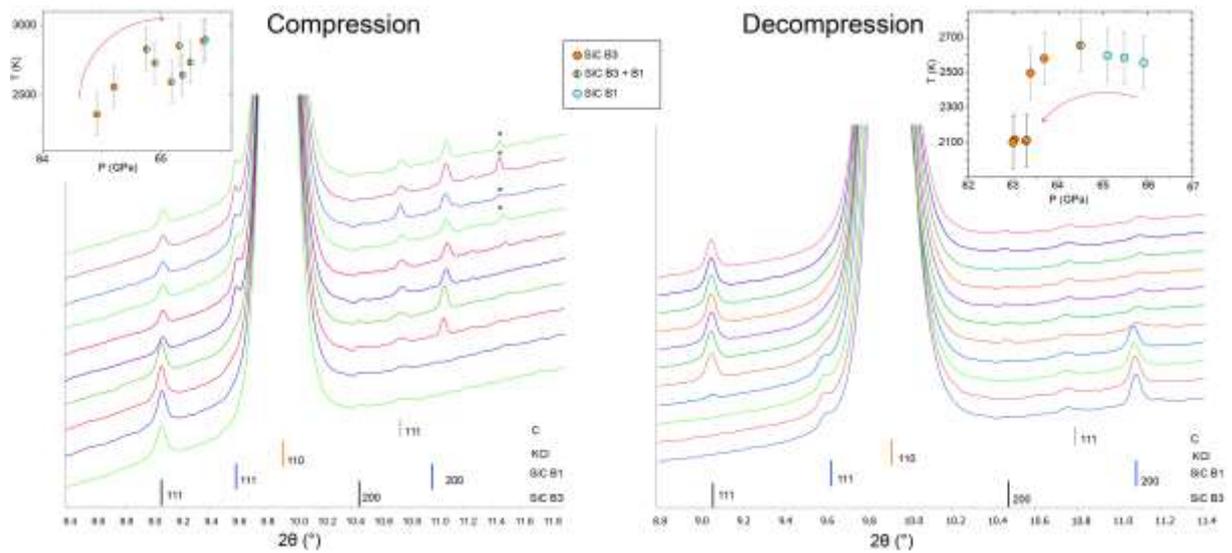

Figure 1

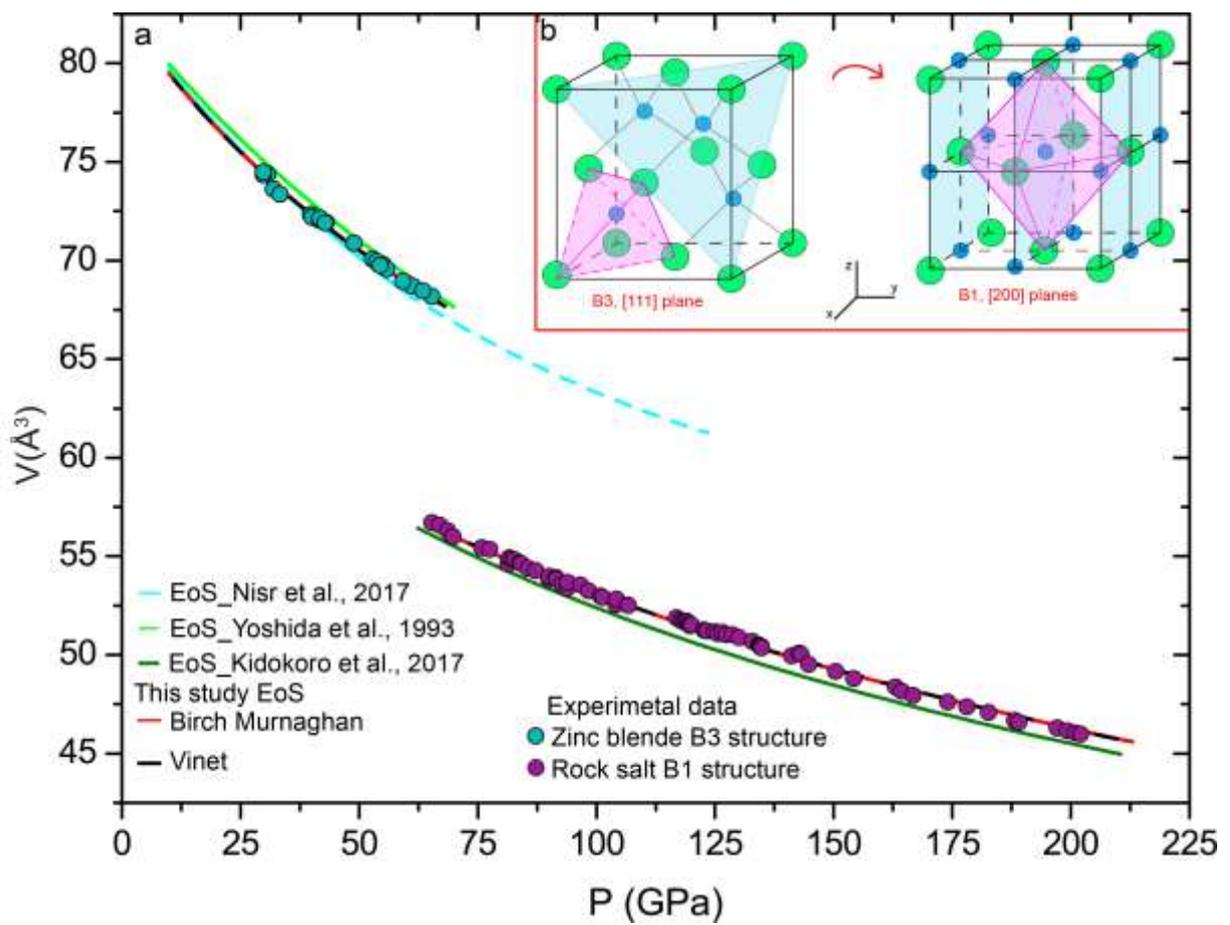

Figure 2

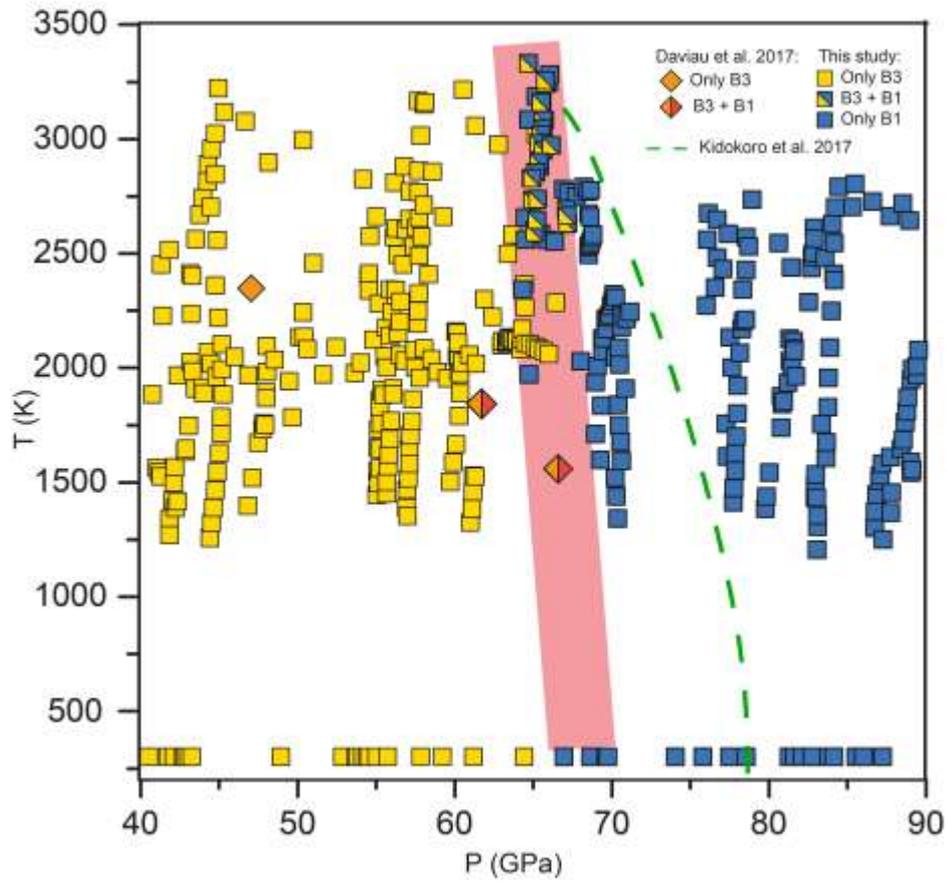

Figure 3

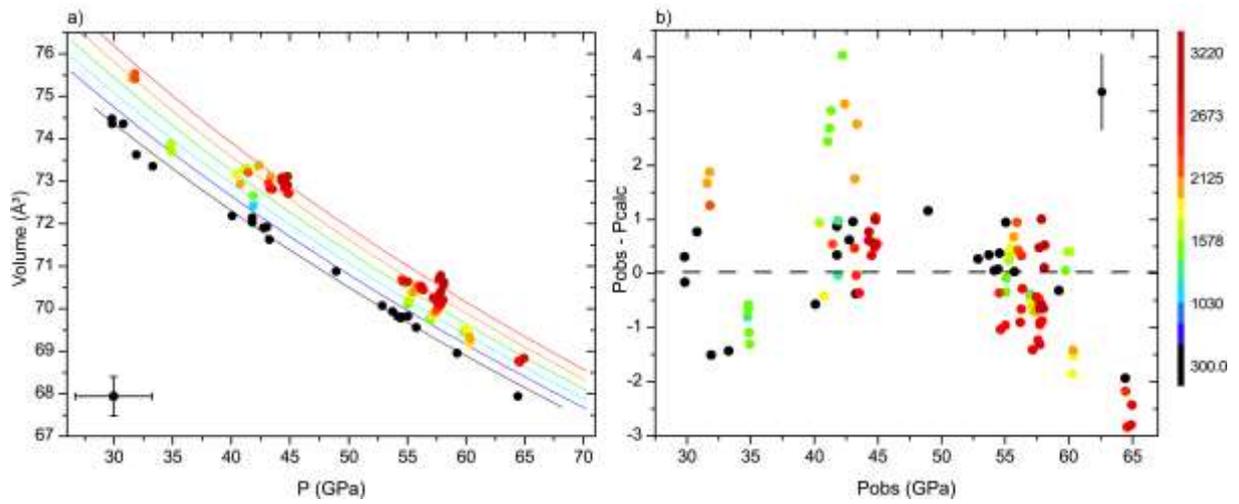

Figure 4

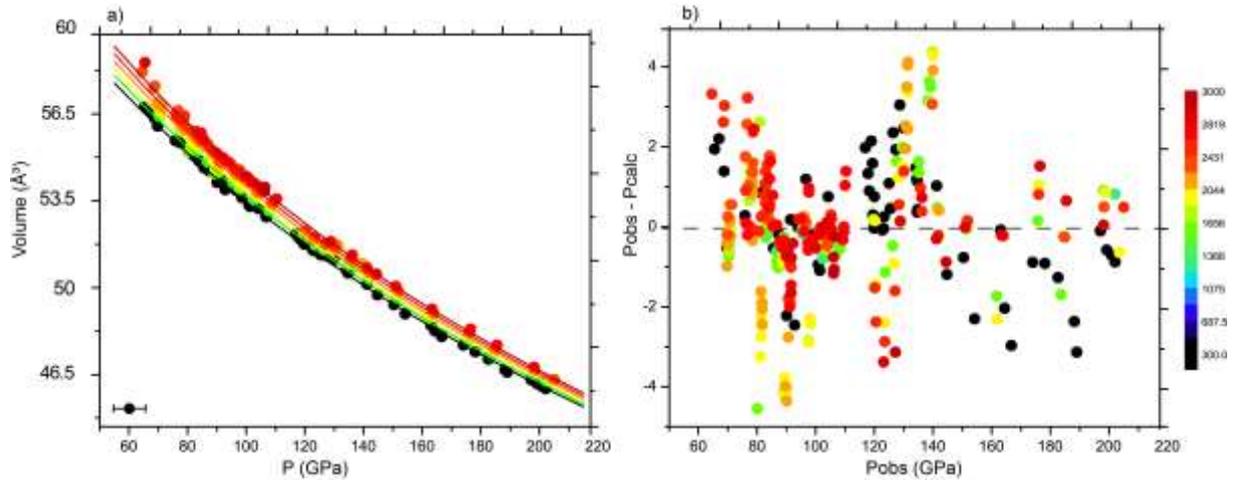

Figure 5

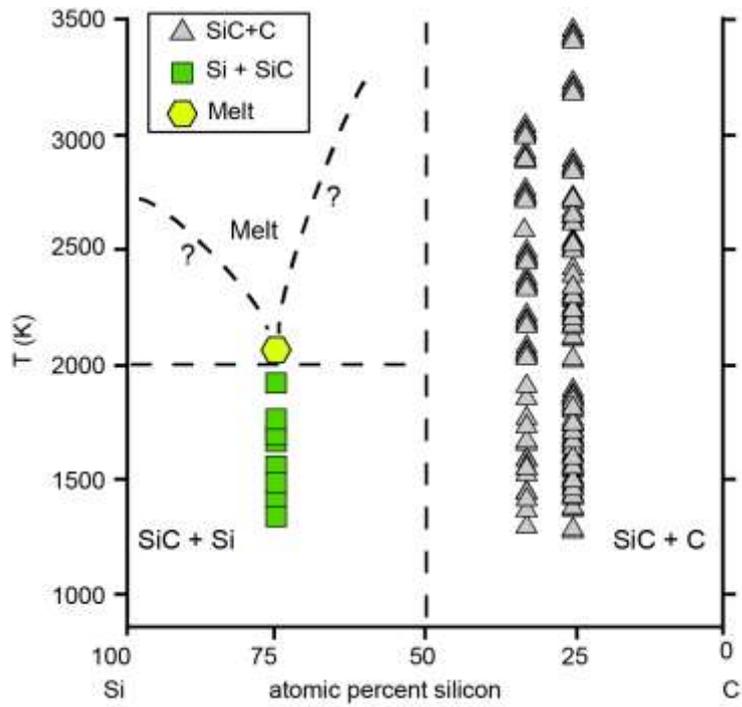

Figure 6

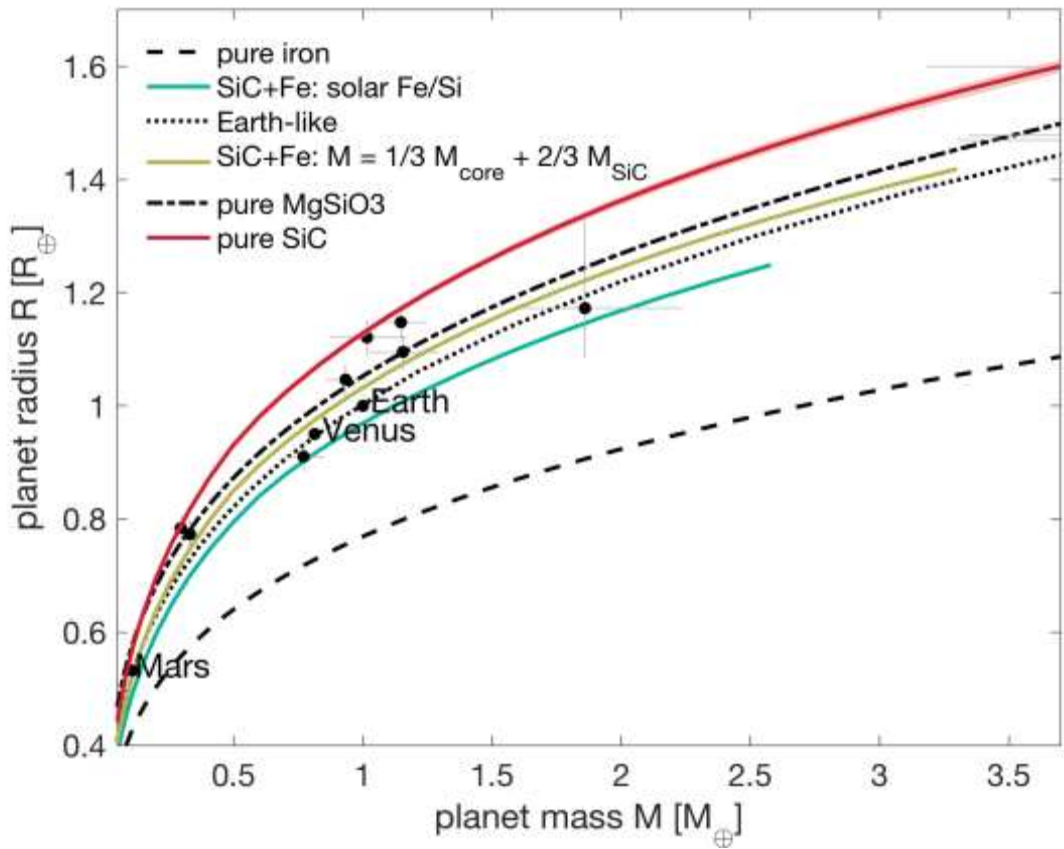

Figure 7

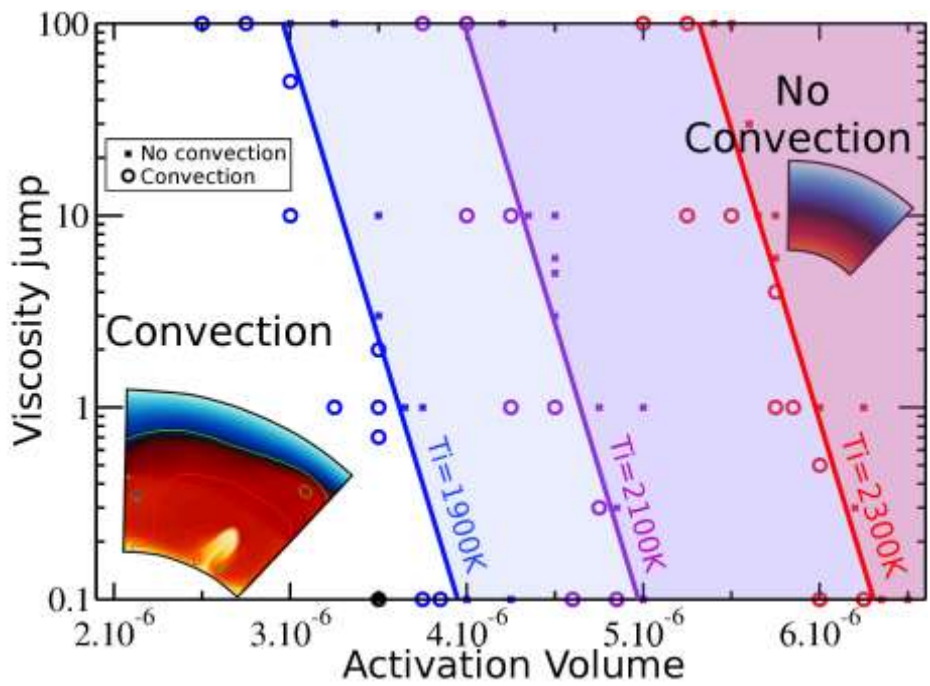

Figure 8